\documentclass[twocolumn]{aastex62}
\usepackage{amsmath}     
\usepackage{wasysym}           
\usepackage{graphicx}
\usepackage{amssymb}
\usepackage{epstopdf}
\usepackage{mathrsfs}
\usepackage{anyfontsize}
\usepackage{natbib}
\usepackage{color}
\usepackage{lipsum}
\usepackage{diagbox}

\shorttitle{The Persistence of Pancakes}
\shortauthors{Coughlin et al.}
\begin{document}
\title{The Persistence of Pancakes and the Revival of Self-gravity in Tidal Disruption Events}
\correspondingauthor{Eric R.~Coughlin}

\author[0000-0003-3765-6401]{Eric R.~Coughlin}
\email{ecoughli@syr.edu}
\affiliation{Department of Physics, Syracuse University, Syracuse, NY 13244, USA}
\affiliation{Department of Astrophysical Sciences, Princeton University, Princeton, NJ 08544, USA}

\author[0000-0002-2137-4146]{C.~J.~Nixon}
\affiliation{School of Physics and Astronomy, University of Leicester, Leicester, LE1 7RH, UK}

\author[0000-0003-1354-1984]{Patrick R. Miles}
\affiliation{Department of Physics, Syracuse University, Syracuse, NY 13244, USA}

\begin{abstract}
The destruction of a star by the tides of a supermassive black hole (SMBH) powers a bright accretion flare, and the theoretical modeling of such tidal disruption events (TDEs) can provide a direct means of inferring SMBH properties from observations. Previously it has been shown that TDEs with $\beta = r_{\rm t}/r_{\rm p} = 1$, where $r_{\rm t}$ is the tidal disruption radius and $r_{\rm p}$ is the pericenter distance of the star, form an in-plane caustic, or ``pancake,'' where the tidally disrupted debris is compressed into a one-dimensional line within the orbital plane of the star. Here we show that this result applies generally to all TDEs for which the star is fully disrupted, i.e., that satisfy $\beta \gtrsim 1$. We show that the location of this caustic is always outside of the tidal disruption radius of the star and the compression of the gas near the caustic is at most mildly supersonic, which results in an adiabatic increase in the gas density above the tidal density of the black hole. As such, this in-plane pancake revitalizes the influence of self-gravity even for large $\beta$, in agreement with recent simulations. This finding suggests that for all TDEs in which the star is fully disrupted, self-gravity is revived post-pericenter, keeps the stream of debris narrowly confined in its transverse directions, and renders the debris prone to gravitational instability.
\end{abstract}


\section{Introduction}
The tidal disruption of a star by a supermassive black hole (SMBH) can illuminate the center of a galaxy for months to years (e.g., \citealt{rees88}), and these tidal disruption events (TDEs) have been discovered with ever-increasing frequency (e.g., \citealt{gezari12, hung17, holoien19, vanvelzen19, gomez20,holoien20}). The number of observed TDEs will rise unprecedentedly in the era of the Vera Rubin Telescope \citep{ivezic19}, and our understanding of the physical evolution of TDEs is necessary for maximizing the potential of such observations.

To this end, three-dimensional hydrodynamical simulations have proved useful for detailing the intricacies of TDEs (e.g., \citealt{bicknell83, evans89, laguna93, lodato09, guillochon13, hayasaki13,coughlin15,mainetti17,golightly19a, miles20, law-smith20}). Alternatively, an analytic approach to describing the evolution of the debris from a TDE is the impulse, or frozen-in, approximation, which assumes that the gas parcels comprising the disrupted star move precisely with the center of mass (COM) until reaching the tidal radius $r_{\rm t}$ and execute ballistic orbits thereafter (e.g., \citealt{lacy82, lodato09, stone13,coughlin19}). The tidal radius $r_{\rm t} = R_{\star}\left(M_{\bullet}/M_{\star}\right)^{1/3}$ is roughly the distance from the SMBH at which the tidal force equals the self-gravity of a star with mass $M_{\star}$ and radius $R_{\star}$. While the frozen-in approximation certainly misses the level of detail captured by numerical simulations, it is relatively simple and unfettered by the enormous range of spatial and temporal scales endemic to TDEs that make hydrodynamical simulations expensive.

For an encounter in which the pericenter distance of the COM, $r_{\rm p}$, is less than the tidal radius, the impact parameter $\beta \equiv r_{\rm t}/r_{\rm p}$ satisfies $\beta > 1$; in this case neglecting the effects of pressure and self-gravity is likely to be upheld reasonably well while the COM is within the tidal radius. However, the neglect of such terms becomes questionable once the stellar debris recedes beyond $r_{\rm t}$. \citet{kochanek94} argued that, if the tidally disrupted debris is freely expanding post-pericenter, then there is a critical $\beta$ above which self-gravity is never important. In terms of the central density of the star $\rho_{\rm c}$ and the average stellar density $\rho_{\star}$, this critical $\beta$ is $\beta \simeq \left(\rho_{\rm c}/\rho_{\star}\right)^{1/3}$ \citep{steinberg19}. For a $\gamma = 5/3$ polytropic star, $\rho_{\rm c}/\rho_{\star} \simeq 8$, and self-gravity should be negligible for $\beta \gtrsim 2$. 

In contrast to this expectation, \citet{steinberg19} found numerically that disruptions of $\gamma = 5/3$ polytropes with $\beta$'s as large as 7, while dominated by the tidal shear of the black hole within a substantial fraction of the tidal radius (and thus validating the neglect of self-gravity that underlies the impulse approximation; see their Figure 3), yielded a large amount ($\gtrsim 30\%$ of the stellar mass) of self-gravitating material by the time the COM exited the tidal radius. \citet{steinberg19} suggested that the origin of this behavior arises from the fact that the gas parcels near pericenter do not necessarily undergo free expansion, but instead can be \emph{compressed} within the plane of the orbit, as found by \citet{coughlin16} for the case of $\beta = 1$. In particular, \citet{coughlin16} showed that the initial conditions inherent to the impulse approximation for a $\beta = 1$ encounter result in the formation of a caustic, or ``pancake,'' within the orbital plane of the stellar COM where the gas parcels would -- in the absence of pressure -- geometrically focus to a line. As such, self-gravity can be ``revived'' at a later time from these compressive effects, even though it may be completely overwhelmed by the tidal shear of the SMBH initially.

Here we show that the pancake described in \citet{coughlin16} exists for large $\beta$'s, supporting the interpretation in \citet{steinberg19}. In Section \ref{sec:model} we describe the model and present results, and we discuss and conclude in Section \ref{sec:discussion}.

\section{In-plane caustic}
\label{sec:model}
\begin{figure*}[htbp] 
   \centering
   \includegraphics[width=0.995\textwidth]{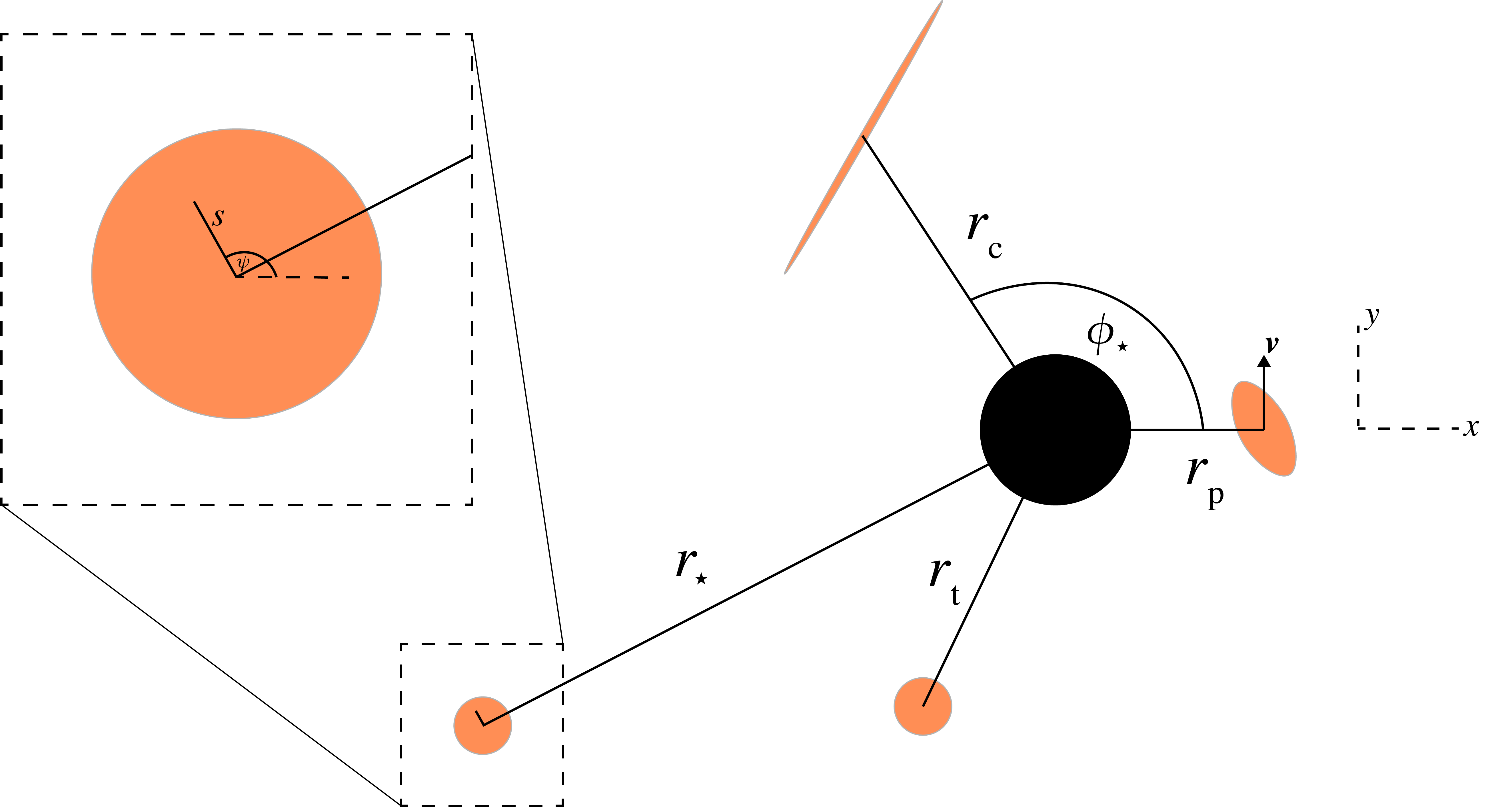} 
   \caption{A diagram of a TDE under the impulse approximation. The distance of the COM from the SMBH, $r_{\star}$, reaches the tidal radius, $r_{\rm t}$, which marks the point where the star is ``destroyed'' and fluid elements follow ballistic orbits thereafter. The orientation of the pericenter distance of the star, $r_{\rm p}$, defines the $x$-axis, while the velocity of the COM at $r_{\rm p}$ defines the $y$-axis. The caustic occurs at $r_{\rm c}$ when the gas parcels in the orbital plane of the stellar COM collapse to a line. The zoom-in of the star in the left of the figure shows the distance between a fluid element and the COM, $s$, and the angle the fluid element makes with the $x$-axis, $\psi$.}
   \label{fig:tde}
\end{figure*}

We assume that the fluid elements of the star move ballistically in the gravitational field of the SMBH once the COM crosses the tidal radius, and that the distance between a fluid element and the COM, $s$, is much less than the distance between the COM and the SMBH, $r_{\star}$. We analyze fluid elements that are within the orbital plane of the COM, which we define as the $x$-$y$ plane with the $x$-direction parallel to the pericenter vector of the COM and $y$ parallel to the velocity vector of the star at pericenter (see Figure \ref{fig:tde}). 

One method of determining the location of the in-plane caustic (the ``pancake'') is to numerically integrate the equations of motion for a large number of fluid elements and find where they cross, which was done in \citet{coughlin16}. However, a more elegant approach is to use the tidal approximation to first simplify the equations of motion, as described in \citet{sari10} and implemented in the case of a TDE in \citet{stone13}. In this approximation the leading order (in $s/r_{\star}$) Lagrangian of a fluid element is

\begin{equation}
\mathscr{L} = 
\dot{s}^2+s^2\dot\psi^2-s^2\left(1+\cos\phi_{\star}\right)^3\left(1-3\cos^2\left[\psi-\phi_{\star}\right]\right) \label{Lag1}.
\end{equation}
Here $\psi$ is the angular position of the fluid element relative to the positive $x$-axis, and we let the stellar COM follow a parabolic trajectory so that $r_{\star}$ and the angle that the COM makes relative to the positive $x$-axis $\phi_{\star}$ satisfy

\begin{equation}
r_{\star} = \frac{2 r_{\rm p}}{1+\cos\phi_{\star}}, \quad \frac{1}{\left(1+\cos\phi_{\star}\right)^2}\frac{d\phi_{\star}}{d\tau} = 1. \label{com1}
\end{equation}
The pericenter distance of the star is $r_{\rm p}$, and the dimensionless, time-like variable $\tau$ is related to time $t$ via $\tau = t\sqrt{GM/(8 r_{\rm p}^3)}$; dots in Equation \eqref{Lag1} denote differentiation with respect to $\tau$. Since the initial position of the COM coincides with the tidal radius, the initial condition for the angle $\phi_{\star}$ satisfies 

\begin{equation}
\cos\left[\phi_{\star}(\tau = 0)\right] = \frac{2}{\beta}-1,
\end{equation}
where $\beta = r_{\rm t}/r_{\rm p}$. 

The equations of motion are the Euler-Lagrange equations, $d/d\tau\left[\partial \mathscr{L}/\partial \dot{s}\right]-\partial\mathscr{L}/\partial s = 0$ and similarly with $s \rightarrow \psi$. Since the entire star moves with the COM upon entering the tidal radius, we have $\dot{s}(\tau = 0) = \dot{\psi}(\tau = 0) = 0$, while the initial position satisfies\footnote{Every term in the Lagrangian is proportional to $s^2$, and hence we can set $s(\tau = 0) = 1$ without loss of generality.} $s(\tau = 0) = 1$ and $\psi(\tau = 0) = \psi_{0}$. The caustic is the location where curves with different $\psi_0$ converge to a single value of $\psi$. As described in \citet{sari10} and \citet{stone13}, there are analytic solutions to the Euler-Lagrange equations, but we find that numerically integrating the equations as a function of the initial angle $\psi_{0}$ offers a straightforward means of finding the pancake.

\begin{figure*}[htbp] 
   \centering
   \includegraphics[width=0.495\textwidth]{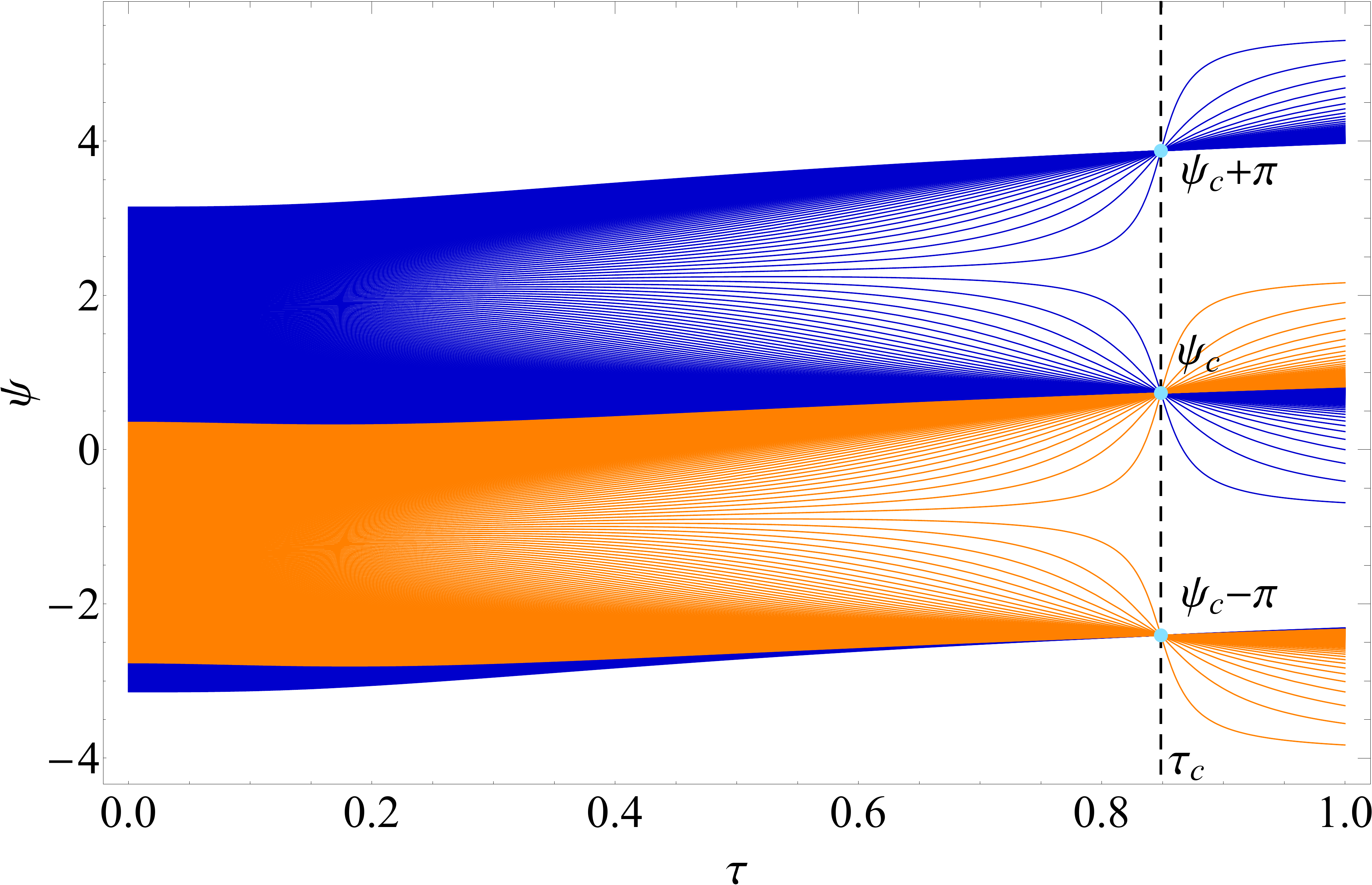} 
    \includegraphics[width=0.495\textwidth]{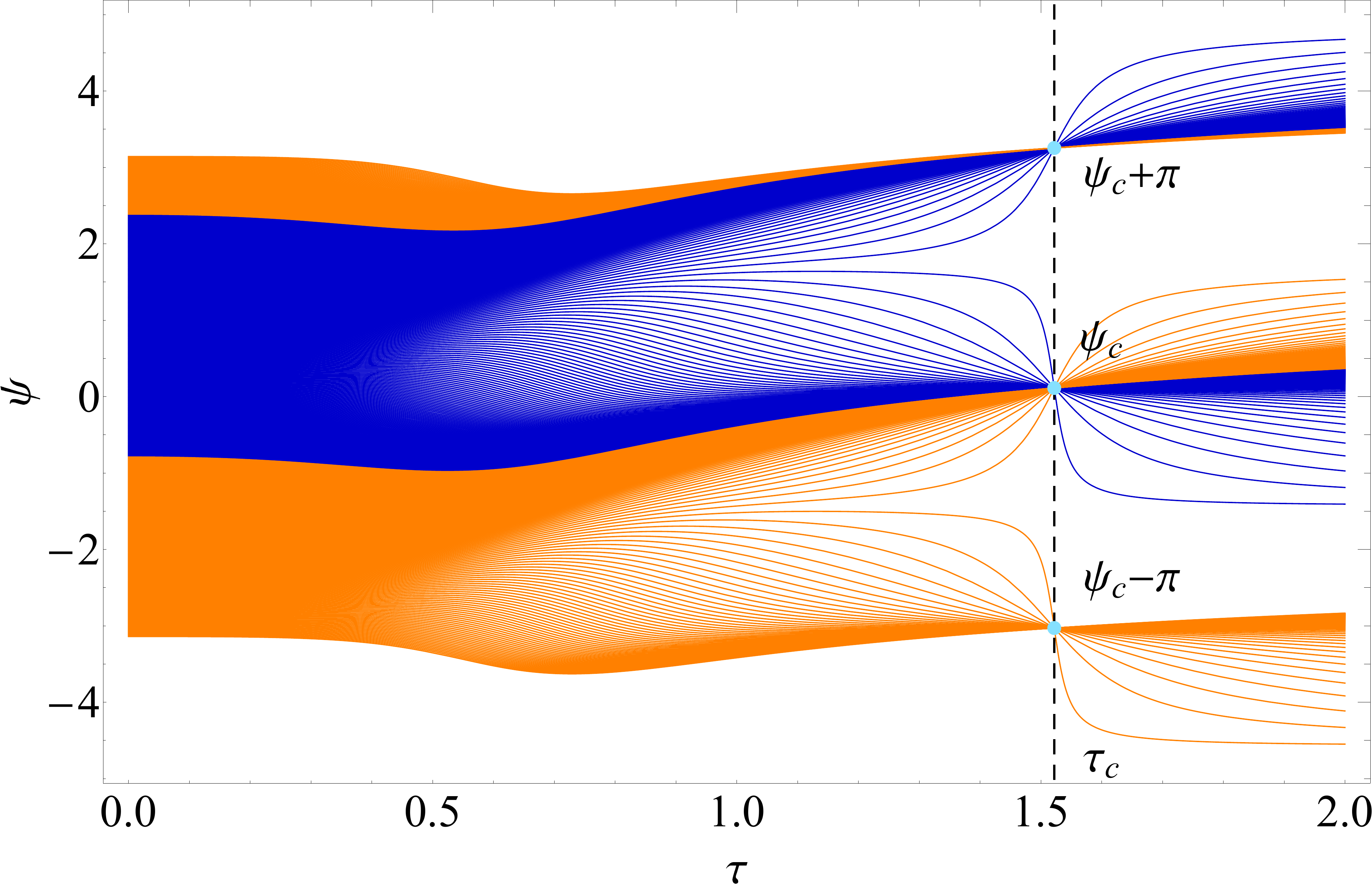} 
    \includegraphics[width=0.4875\textwidth]{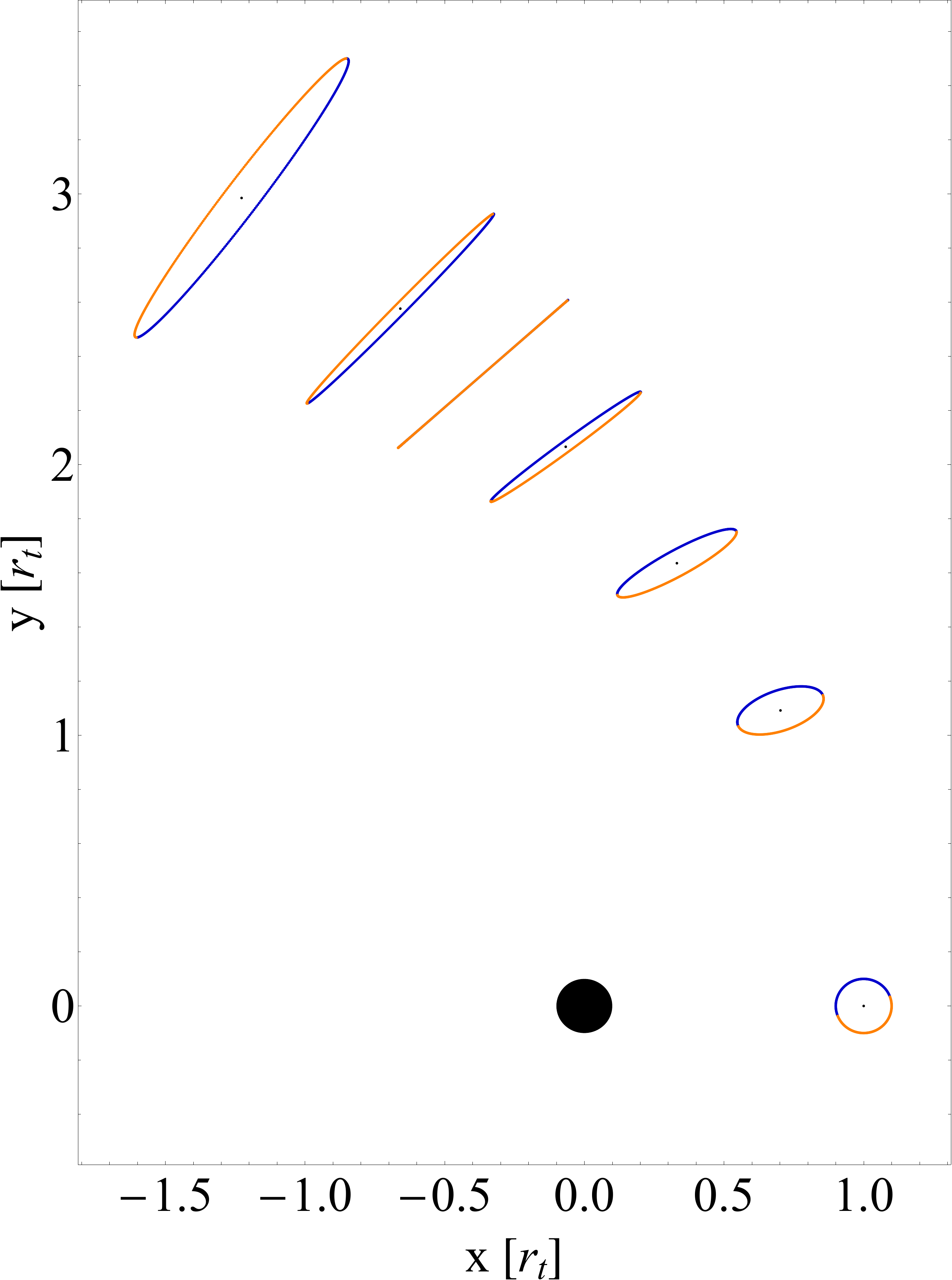} 
    \includegraphics[width=0.505\textwidth]{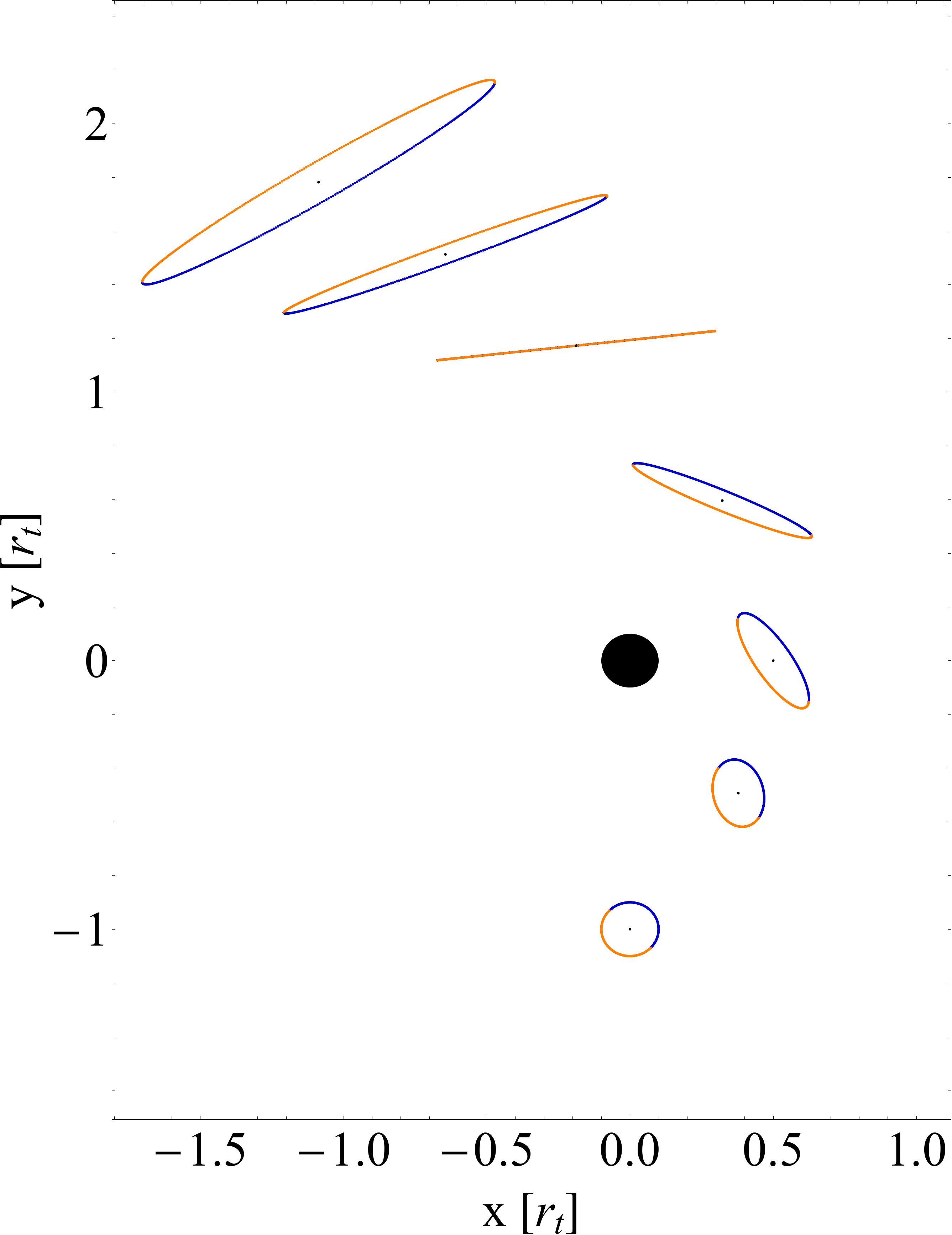} 
   \caption{The top two panels show the evolution of the angle $\psi$ as a function of time for a number of different initial angles, which sample the range $\{-\pi$, $\pi\}$; the top-left (top-right) panel is for $\beta = 1$ ($\beta = 2$). The vertical, dashed lines indicate the time at which the caustic occurs $\tau_{\rm c}$, which is where all of the curves intersect at $\psi_{\rm c}$ or $\psi_{\rm c}\pm \pi$, these values denoted by the cyan points.  At this time, therefore, every circle of points within the plane of the disrupted star collapses to a line, and the front and back of the star switch places; the colors denote the front (blue) and back (orange) of the star immediately prior to the caustic. The bottom two panels illustrate the positions of the gas parcels when $\beta = 1$ (left) and $\beta = 2$ (right) at times that bracket the pre and post-caustic evolution. The color-coding matches that of the lines in the top two panels.} 
   \label{fig:psicurves}
\end{figure*}

The top two panels of Figure \ref{fig:psicurves} show the evolution of the angle $\psi(\tau)$ for a number of different $\psi_{0}$ when $\beta = 1$ (left panel) and $\beta = 2$ (right panel). For $\beta = 1$ ($\beta = 2$), at a time of $\tau_{\rm c} \simeq 0.849$ ($\tau_{\rm c} \simeq 1.52$) all of the curves intersect at the common angle $\psi_{\rm c} \simeq 0.732$ ($\psi_{\rm c} = 0.112$) or $\psi_{\rm c}\pm \pi$, shown by the cyan points. Thus, at $\tau_{\rm c}$ the orbits of the fluid elements collapse to a line, and the front and back of the star switch places; the curves are color-coded according to the front (blue) and back (orange) immediately prior to the caustic\footnote{The ``front of the star'' is the collection of gas parcels within $\psi_{\rm c}$ and $\psi_{\rm c}+\pi$ at a time just prior to $\tau_{\rm c}$.}. The bottom two panels show a ring of fluid elements for $\beta = 1$ (left) and $\beta = 2$ (right) at a number of different times to illustrate the formation of the caustic. 

\begin{figure}[htbp] 
   \centering
   \includegraphics[width=0.475\textwidth]{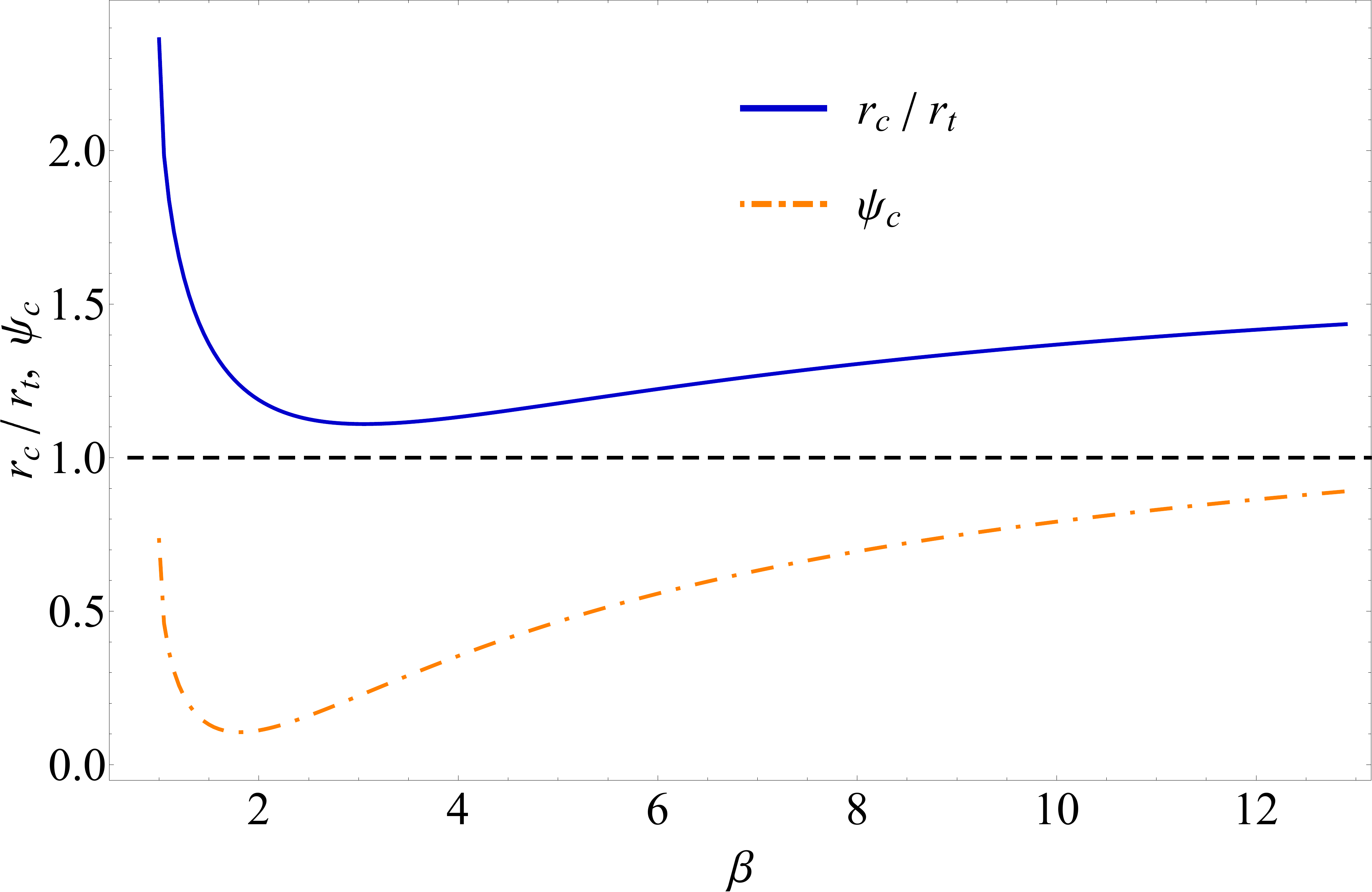} 
   \caption{The position of the COM normalized by the tidal radius, $r_{\rm c}/r_{\rm t}$ (blue), and the angle that the caustic makes, $\psi_{\rm c}$ (orange, dot-dashed), at the time the caustic occurs as functions of $\beta = r_{\rm t}/r_{\rm p}$. The blue curve shows that the caustic exists independent of $\beta$ and outside of the tidal disruption radius.}
   \label{fig:rcbeta}
\end{figure}

Figure \ref{fig:rcbeta} shows the position of the COM when the caustic occurs, $r_{\rm c}$, divided by the tidal radius, $r_{\rm t}$, as a function of $\beta$ (blue curve) and the angle that the caustic makes with the positive $x$-axis (orange, dot-dashed curve). We see that the distance where the caustic occurs reaches a relative minimum from the SMBH of $\simeq 1.1 r_{\rm t}$ at $\beta \simeq 3$ and the caustic is always outside of the tidal radius. The angle that the caustic makes with the $x$-axis also reaches a minimum of $\psi_{\rm c} \simeq 0.1$ near $\beta \simeq 2$, where the caustic is nearly parallel to the pericenter vector (see the bottom-right panel of Figure \ref{fig:psicurves}). 

\begin{figure}[htbp] 
   \centering
   \includegraphics[width=0.47\textwidth]{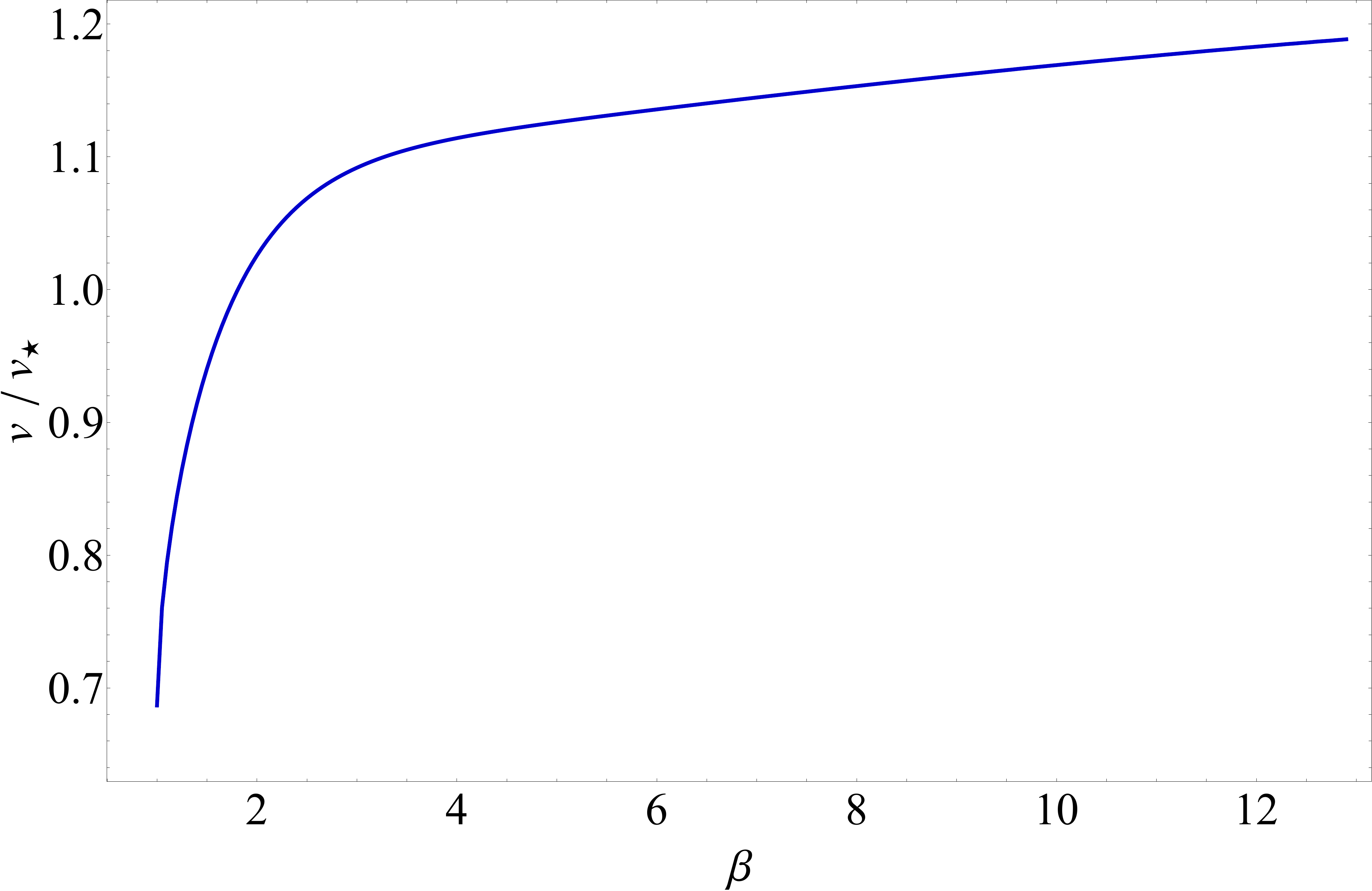} 
   \caption{The speed at which material converges onto the COM normalized by the stellar escape speed, $v / v_{\star}$, as a function of $\beta$. Because this ratio is always on the order of unity, the compression that occurs at the in-plane pancake is at most very mildly supersonic, meaning that a strong shock does not form.}
   \label{fig:vonvstar}
\end{figure}

Figure \ref{fig:psicurves} shows that there are regions near the angles $\psi_{\rm c}\pm \pi/2$ at $\tau_{\rm c}$ where fluid elements rapidly increase or decrease by $\pi$. The fluid elements at $\psi_{\rm c}\pm \pi/2$ are thus directly in front of ($\psi_{\rm c}+\pi/2$) and behind ($\psi_{\rm c}-\pi/2$) the COM at the time the caustic occurs and reach a coordinate singularity $s(\tau_{\rm c}) = 0$. At $\tau_{\rm c}-\epsilon$ these fluid elements move only in the $s$-direction in the limit that $\epsilon \rightarrow 0$, and hence the rate at which the surface of the debris converges toward the COM (i.e., the rate of change of the stream diameter) at $\tau_{\rm c}$ is $2|\partial s/\partial t| = \beta^{3/2}v_{\star}|\partial s/\partial \tau|/2$, where $v_{\star} = \sqrt{2GM_{\star}/R_{\star}}$ is the escape speed of the star with mass $M_{\star}$ and radius $R_{\star}$ and the derivative is taken as $\epsilon \rightarrow 0$. Figure \ref{fig:vonvstar} shows the ratio $v / v_{\star}$ as a function of $\beta$, and illustrates that this value is of the order unity. Thus the compression from the in-plane caustic is approximately adiabatic and does not form a strong shock.

\begin{figure}[htbp]
   \centering
   \includegraphics[width=0.47\textwidth]{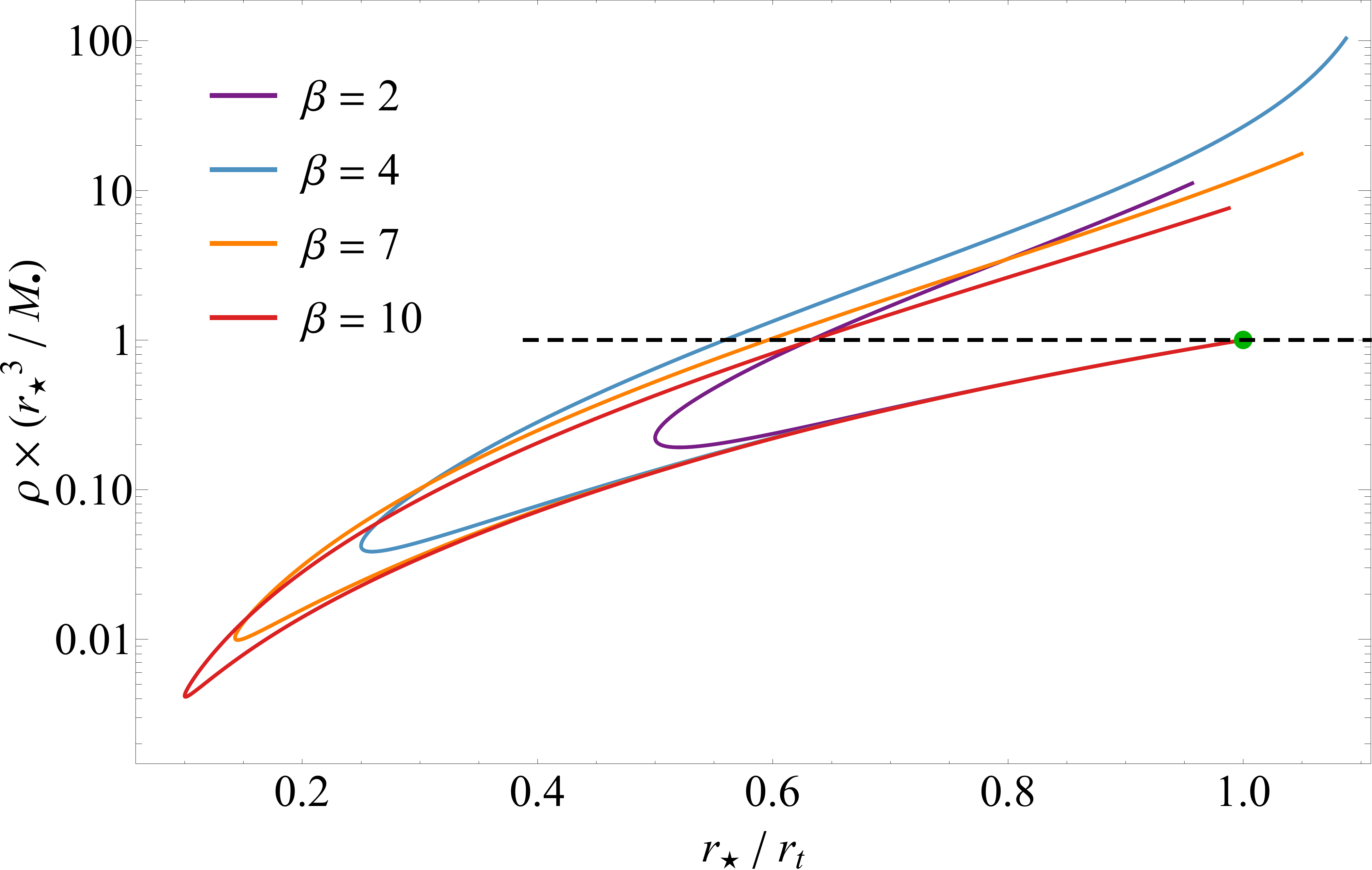} 
   \caption{The ratio of the average density of a ring of fluid elements $\rho$ to the density of the black hole, $\rho_{\bullet} \simeq M_{\bullet}/r_{\star}^3$; this ratio has been normalized to unity at the time the star enters the tidal radius of the SMBH, that time shown by the green circle. This ratio is plotted as a function of the distance of the stellar COM relative to the tidal radius, and different curves correspond to the $\beta$ shown in the legend. In time each curve starts at the green point, the center of mass moves to smaller radii (left), reaches pericenter, and moves back to larger radii (right). Because of the compressive effects of the caustic, the ratio of the density to the black hole density increases post-pericenter, and would diverge at the location of the caustic in the absence of pressure.}
   \label{fig:rhorhob}
\end{figure}

By definition the mass of a fluid element is conserved, and hence the density\footnote{The out-of-plane motion of the fluid elements, which we ignore here, implies that the true variation in the density is more complicated than this; we discuss this further in Section \ref{sec:discussion} below, but the solution for the density given by Equation \eqref{rho} can be more accurately interpreted as a column density, i.e., the true density integrated over the height of the debris.} $\rho$ satisfies

\begin{equation}
\rho(s,\psi,\tau) = \rho_0(s_0,\psi_0)J^{-1}, \label{rho}
\end{equation}
where $J$ is the Jacobian that relates the time-dependent positions of the fluid elements $\{s,\psi\}$ to their initial positions $\{s_0,\psi_0\}$. Since the equations of motion are independent of $s_0$ the Jacobian is simply $J = \partial\psi/\partial\psi_0$. Figure \ref{fig:rhorhob} shows the average density of a ring of fluid elements as a function of the position of the center of mass relative to the tidal radius for the $\beta$ in the legend; the density was determined by interpolating the Jacobian for a number of different $\psi_0$, calculating the derivative, and averaging over the particles. The density is plotted relative to the tidal density of the SMBH, $\rho_{\bullet} \simeq M_{\bullet}/r_{\star}^3$, and this ratio is normalized to unity when the star enters the tidal radius, which reflects the fact that the self-gravity of the material is dominated by the tidal field of the black hole interior to $\sim r_{\rm t}$. As functions of time all of the curves start at the green point, move to the left where they reach their pericenter distance ($ = 1/\beta$), and move back to larger distances from the SMBH. Prior to hitting the caustic the density increases back above the tidal density of the SMBH owing to the compression in the plane.

\section{Discussion and Conclusions}
\label{sec:discussion}
The analysis in Section \ref{sec:model} indicates that a ``pancake,'' or a caustic where the fluid elements within the orbital plane of a tidally disrupted star collapse to a line, exists for TDEs with large $\beta$ (Figure \ref{fig:rcbeta}). The rate of compression of the gas near the caustic is at most mildly supersonic (Figure \ref{fig:vonvstar}), which implies that the convergence does not generate a strong shock. The ratio of the density to the tidal density of the SMBH increases as the stellar COM moves out from pericenter (Figure \ref{fig:rhorhob}), and would diverge to infinity at the location of the caustic in the absence of pressure.

Gas pressure maintains a finite maximum in the density, which we can estimate by equating the pressure $p$ to the ram pressure of the converging flow $\rho v^2$. Since $v / v_{\star} \simeq 1$ (Figure \ref{fig:vonvstar}) with $v_{\star} = \sqrt{2GM_{\star}/R_{\star}}$, the compression occurs adiabatically and the specific entropy is approximately preserved, and the sound speed increases to roughly the stellar escape speed. It follows that at the time of maximum compression the density is comparable to the original stellar density, because the configuration reaches a state with zero velocity and the sound speed is comparable to the escape speed of the star (i.e., it is in the same equilibrium state as the original star). Since the caustic occurs outside of the tidal disruption radius, this demonstrates that the in-plane pancake revives self-gravity in high-$\beta$ encounters, in agreement with the findings and interpretation of \citet{steinberg19}. 

More quantitatively, the velocity profile is nearly homologous in the direction perpendicular to the caustic as the flow converges, which arises from the independence of the equations of motion on the initial radius within the star. Approximating the distribution of debris as a cylinder, which is reasonable from the bottom panels of Figure \ref{fig:psicurves}, then the convergence will excite predominantly the $f$-mode of the adiabatic cylinder. We therefore expect the pancake to generate time-dependent oscillations of the fluid at the frequency associated with the $f$-mode of an adiabatic cylinder, in agreement with the findings of \citet[see their Figure 4]{coughlin20b}. 

The caustic augments the self-gravity of the material to the point where the stretching stream of debris becomes quasi-hydrostatic in the transverse directions. As such, the stream is gravitationally unstable provided that the dominant contribution to the hydrostatic balance comes from gas pressure, with the instability leading to the formation of self-bound knots that are distributed along the length of the stream in a manner that can be determined from a stability analysis \citep{coughlin15, coughlin20a}. The fact that the in-plane pancake exists for large $\beta$ implies not only that self-gravity remains important for confining the stream in these deeply plunging encounters, but that variability in the fallback rate as distinct clumps of debris return to pericenter is a feature of TDEs irrespective of how large $\beta$ becomes. 

We did not consider the motion perpendicular to the orbital plane of the star, which generates a distinct caustic as fluid elements cross the plane of the COM; as $\beta$ becomes large, this caustic occurs roughly coincidently with the pericenter distance of the star \citep{carter83, stone13}. The motion of the gas out of the plane will certainly affect the estimates of the density. However, the out-of-plane motion should be nearly symmetric about the point of maximum compression (the ``bounce''; \citealt{carter83, stone13}), and hence the density will increase (decrease) more rapidly as the star approaches (recedes from) pericenter. We therefore expect the net effect of the out-of-plane motion to be small by the time the in-plane caustic occurs, and hence our estimate of the density near the caustic -- that it is comparable to the original stellar density -- is likely unaffected by these additional physical considerations. The simulations of \citet{steinberg19} point to the validity of these arguments.

The arguments above suggest that the in-plane compression of the gas augments the average density of the material to a value comparable to that of the initial star. Because of the fact that the initial star possesses a density profile that has regions above and below the average density, we expect the same to be true of the post-pancake debris: the highest-density regions near the COM will be safely above the self-gravitating limit, while the low-density, outer extremities will likely not be self-gravitating. There are also subtleties related to the breakdown of the approximation that the entire star moves with the center of mass at the tidal radius; in reality different shells of the star likely possess varying degrees of differential motion with respect to one another that consequently modify the nature of the caustic (see Figure 10 of \citet{coughlin16} and their discussion related to this point). Again, the simulations of \citet{steinberg19} reflect the notion that these additional complications do not completely stifle the revitalizing influence of self-gravity (those authors also note that the self-gravitating material is confined to a denser ``core'' of material surrounded by a non-self-gravitating ``sheath''). 

We adopted the tidal limit in our analysis in Section \ref{sec:model}, which assumes that the distance of the stellar COM to the SMBH is much greater than the size of the star itself, and we also did not include general relativistic effects. These approximations break down at sufficiently large $\beta$, the former (latter) becoming increasingly important as the SMBH mass decreases (increases; e.g., \citealt{kesden12, stone13, gafton15, stone19, darbha19}). It would be interesting to redo the analysis with general relativistic terms included to determine the modifications to the location at which the in-plane pancake occurs or, indeed, if it precludes its existence altogether above some $\beta$.

Another consequence of the vertical motion (not included in our model) arises from the increase in the gas pressure as a result of the compression near pericenter, with the increase either occurring adiabatically or through the formation of a shock \citep{bicknell83, kobayashi04, brassart08, guillochon09}. It is possible that, at the expense of reducing the vertical motion of the gas, the increase in the pressure serves to impart a more isotropic rebound of the fluid, altering the in-plane motion and correspondingly the location (or existence) of the pancake. While such a redistribution of the kinetic energy probably occurs to some degree, this effect likely does not modify the motion of the gas within the plane to the point where the in-plane pancake can be prevented entirely, because the caustic generated from the vertical collapse occurs simultaneously only for a ring of fluid elements (out of the plane) at a given initial displacement from the center of mass. Instead of compressing isotropically around the center of mass, the points of maximum compression at a given time (i.e., where the caustic occurs) therefore coincide with a line within the orbital plane. The pressure \emph{gradient} responsible for accelerating the fluid is thus still maximized out of the plane, and hence the pressure that builds to resist the vertical compression serves primarily to halt (and reverse) that compression instead of redistributing the energy within the plane. The simulations of \citet{steinberg19} also substantiate the notion that the in-plane pancake is not prevented by this additional effect.

The role of stellar spin in modifying the disruption dynamics has also been investigated relatively recently \citep{golightly19b, kagaya19, sacchi19}. It is straightforward to include this effect in the model presented in Section \ref{sec:model}; if the star is initially rotating with a uniform angular velocity $\Omega_{\star} = \lambda\sqrt{GM_{\star}/R_{\star}^{3/2}}$, where $\lambda$ is the rotational velocity of the star relative to breakup (i.e., for $\lambda = 1$ the star is rotating near breakup and hence we require $\lambda \lesssim 1$), then the only modification\footnote{Assuming that the stellar spin axis is perfectly aligned or anti-aligned with the angular momentum vector of the star; if the spin axis is tilted, then $\lambda$ is the projection of the angular velocity onto the orbital plane of the star (see also \citealt{golightly19b} for further analysis of the tilted case).} to a rotation-less disruption is that the initial angular velocity of a fluid element within the star becomes 

\begin{equation}
\frac{\partial\varphi}{\partial \tau}\bigg{|}_{\tau = 0} = \frac{\lambda\sqrt{8}}{\beta^{3/2}},
\end{equation}
which clearly reduces to the case analyzed in detail in Section \ref{sec:model} when $\lambda = 0$. Note that positive $\lambda$ implies that the star is rotating such that its angular momentum vector is aligned with that of the orbit of the COM, while negative $\lambda$ implies retrograde rotation. What this expression illustrates is that, even for stars rotating near breakup\footnote{Such an encounter would be extremely rare if the disrupted star was placed on a centrophyllic orbit through ordinary two-body relaxation in the host galaxy, but could be more plausible if the star entered the loss cone of the SMBH through a more exotic mechanism, such as a repeated encounter (e.g., \citealt{sacchi19}) or through interactions with a SMBH binary \citep{coughlin17}.}, the initial rotation of the star is negligible once $\beta$ becomes greater than about 2. Thus, while rapid stellar rotation could substantially modify the disruption dynamics for modest $\beta$ (as suggested by the simulations of \citealt{golightly19a}, who found that fractions of breakup $\lambda \gtrsim 0.2$ were necessary to generate even modest differences between irrotational disruptions for $\beta = 1$) and conceivably prevent the formation of the caustic, the existence of the pancake is not heavily affected by the initial spin of the star once $\beta$ becomes large. 

\acknowledgements
We thank Elad Steinberg and the anonymous referee for useful comments and suggestions. E.R.C. acknowledges support from NASA through the Hubble Fellowship Program, grant No.~HST-HF2-51433.001-A awarded by the Space Telescope Science Institute, which is operated by the Association of Universities for Research in Astronomy, Incorporated, under NASA contract NAS5-26555. E.R.C. and P.R.M. acknowledge support from the National Science Foundation through grant AST-2006684. C.J.N. is supported by the Science and Technology Facilities Council (grant number ST/M005917/1). This project has received funding from the European Union’s Horizon 2020 research and innovation program under the Marie Sk\l{}odowska-Curie grant agreement No 823823 (Dustbusters RISE project).

\newpage

\bibliographystyle{aasjournal}

\end{document}